\documentclass{aa} 

\usepackage{graphics}
\begin{document}

\title{Reconstruction of solar activity for the last millennium using $^{10}$Be data
}

\author{I. G. Usoskin\inst{1}, K. Mursula\inst{2}, S. Solanki\inst{3}, M. Sch\"ussler\inst{3}
           \and
           K. Alanko\inst{2}
           }

    \offprints{I.Usoskin}

\institute{Sodankyl\"a Geophysical Observatory (Oulu unit), FIN-90014
     University of Oulu, Finland\\
     email: ilya.usoskin@oulu.fi
     \and
     Department of Physical Sciences, FIN-90014 University of Oulu, Finland
          \and
      Max Planck Institute for Aeronomy, Katlenburg-Lindau, Germany
              }

    \date{Received  .....  ; accepted ....}

\titlerunning{Solar activity: Millennium scale}


\abstract{
In a recent paper (Usoskin et al., 2002a), we have
 reconstructed the concentration of the cosmogenic $^{10}$Be
 isotope in ice cores  from the measured sunspot numbers by
  using physical models for $^{10}$Be production in the Earth's atmosphere,
  cosmic ray transport in the heliosphere, and evolution of the Sun's open magnetic flux.
Here we take the opposite route:
 starting from the $^{10}$Be concentration measured in ice cores from Antarctica and Greenland,
 we invert the models in order to reconstruct the 11-year averaged sunspot
 numbers since 850 AD.
The inversion method is validated by comparing the reconstructed sunspot numbers
 with the directly observed sunspot record since 1610.
The reconstructed sunspot record exhibits a prominent period of about 600 years,
 in agreement with earlier observations based on cosmogenic isotopes.
Also, there is evidence for the century scale Gleissberg cycle and a number of
 shorter quasi-periodicities whose periods seem to fluctuate in the millennium
 time scale.
This invalidates the earlier extrapolation of multi-harmonic representation of sunspot
 activity over extended time intervals.
}

\maketitle

\section{Introduction}
Sunspot numbers (SN) form the most common index of solar activity
 reflecting the varying strength of the hydromagnetic dynamo process which generates the
 solar  magnetic field.

There are two approaches to reconstruct SN
 for times prior to regular direct observations.
The first approach is based on mathematical extrapolation using the statistical
 properties of the SN record observed during the last 300 years.
Such models provide, e.g., a single (11.1-year) carrier frequency
 or a multi-harmonic representation of the measured SN,
 which is then extrapolated backward in time (\cite{scho55,nago97,rigo01}).
Some models also use fragmentary data from naked-eye sunspot observations or sightings of aurorae.
The disadvantage of this approach is that it is not a reconstruction based upon a
description of  physical processes but rather
 a prediction based on extrapolation.
Clearly such models cannot include periods exceeding  the time span of
 observations upon which the extrapolation is based.
Hence, the pre- or post-diction becomes increasingly unreliable with
 extrapolation time
and its accuracy is hard to estimate.

The second approach is based on measured archival proxies of SN such as cosmogenic
 isotopes (\cite{see,ling63,beer90,obri91,damo91,beer00}).
While this approach is based upon real measurements, the quantitative relationship between SN and
 the measured cosmogenic isotope concentration is
 usually described  in an ad-hoc fashion by  a simple inverse relation.

\begin{figure}
\resizebox{\hsize}{!}{\includegraphics{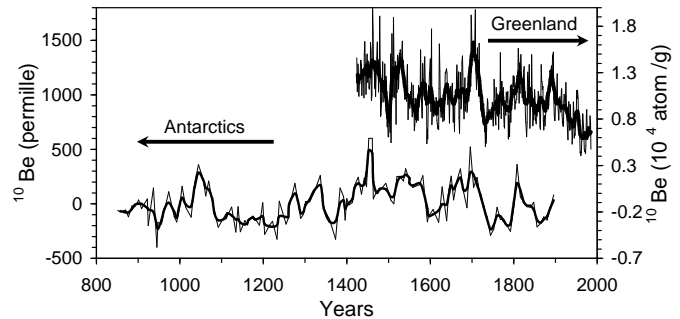}}
     \caption{Raw and smoothed $^{10}$Be data.
     The lower curves (left axis) give the raw (thin curve) and the 1-2-1 filtered 8-year-averaged data
      from Antarctica (\cite{bard97}).
     The upper curves (right axis) show the raw and the 11-year smoothed yearly data from
      Greenland (\cite{beer90}).}
     \label{Fig:data}
\end{figure}

In this paper we follow an improved version of the second approach and reconstruct the SN for the last
 millenium using a physical model with a realistic relation between the concentration of the
 cosmogenic $^{10}$Be isotope in polar ice and the SN.
The $^{10}$Be concentration corresponds well, with a response time of 1-2 years,
 to the flux of primary cosmic rays in the vicinity of the Earth (\cite{bard97,beer00}).
We use two sets of $^{10}$Be data shown in Fig.~\ref{Fig:data}.
One is the annual series of $^{10}$Be concentration in Greenland ice (Dye-3 site, 65.15 N 43.82 W)
 for the years 1424--1985 (\cite{beer90}).
The other series gives the $^{10}$Be concentration with roughly 8-year averaging in Antarctic ice
 (South Pole) for the years 850--1900 (\cite{rais90,bard97}).

\section{Model}

\begin{figure}
\resizebox{\hsize}{!}{\includegraphics{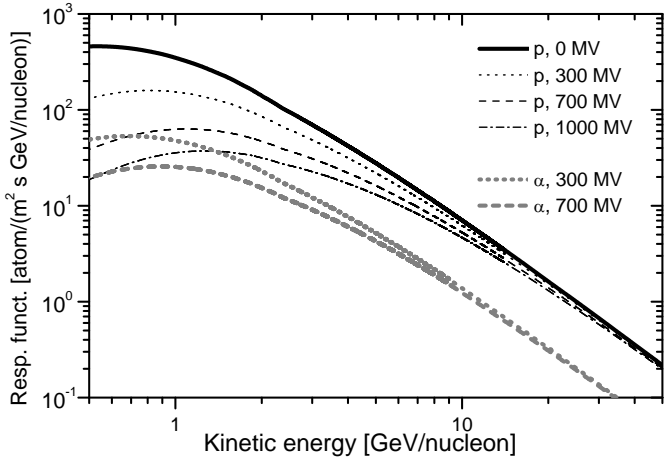}}
     \caption{The differential response function of $^{10}$Be to the two main species
     of galactic cosmic rays, protons and $\alpha-$particles,
      for different values of the heliospheric modulation strength, $\Phi$.}
     \label{Fig:resp_f}
\end{figure}

In a recent paper (\cite{usos02}), we have developed a model which connects the sunspot
 number $N$ with
 the cosmic ray flux (CR) through the source term, $S$, the open solar magnetic flux,
 $F_o$, and the modulation strength, $\Phi$, through a sequence of steps:
\begin{equation}
N\stackrel{(1)}{\longrightarrow} S\stackrel{(2)}{\longrightarrow} F_o\stackrel{(3)}{\longrightarrow}
\Phi\stackrel{(4)}{\longrightarrow}CR\stackrel{(5)}{\longrightarrow}\,^{10}{\rm Be}
\label{Eq:model1}
\end{equation}
Steps (1) and (2) are based upon a model for the open solar magnetic flux (\cite{sola00}),
 steps (3) and (4) on a 1D model of heliospheric transport of CR (\cite{usos02a}).
For step (5) we have earlier (\cite{usos02}) assumed that the $^{10}$Be production
 and the related concentration
 in polar ice is proportional to the flux of CR with an energy of about 2 GeV.
Here we calculate step (5) using a more realistic production rate of $^{10}$Be by CR in
 the Earth's atmosphere.
The local $^{10}$Be production rate, $R$, is given by
\begin{equation}
R=\int_{P_c}^{\infty} X(P)\cdot Y(P)dP,
\label{Eq:prod}
\end{equation}
\begin{figure}
\resizebox{\hsize}{!}{\includegraphics{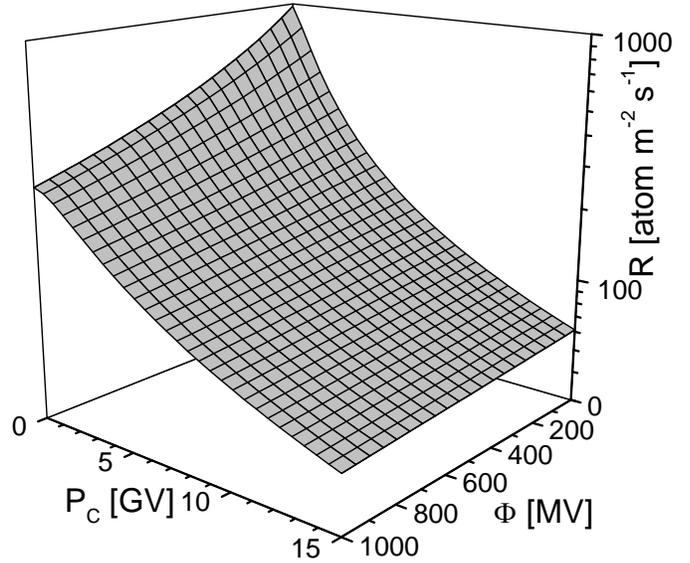}}
     \caption{Production rate of $^{10}$Be in the atmosphere
      as a function of the heliospheric modulation strength, $\Phi$, and
      the local rigidity cutoff, $P_c$.}
     \label{Fig:map_pc}
\end{figure}
where $P_c$, $X(P)$, $Y(P)$, and $P$ are the local rigidity cutoff,
 the differential CR spectrum near the Earth, the differential yield function of $^{10}$Be
 production, and the primary CR particle's rigidity, respectively.

Heavier species (mostly $\alpha-$particles) compose a part of all CR, about 6\%
 in particle number or about 25\% in nucleon number according to recent high-precision
 measurements (\cite{see,boen99,alca00,alca00a}).
Moreover, since heavier nuclei are less modulated in the heliosphere than protons,
 their contribution is larger at lower energies (about 1--2 GeV/nucleon) which are most
 effective for $^{10}$Be production in the atmosphere.
Therefore, we have taken $\alpha-$particles into account in the present study.
Species heavier than $\alpha-$particles can be neglected because of their low abundance in CR).

Using differential CR (protons and $\alpha-$particles) spectra obtained from
 our heliospheric model (\cite{usos02a}),
 and the differential yield function of $^{10}$Be production in the atmosphere
 (\cite{webb03}), we have calculated the
 $^{10}$Be differential response function (Fig.~\ref{Fig:resp_f}) which is the
 integrand appearing in Eq.~(\ref{Eq:prod}).
One can see that the contribution of heavier nuclei to $^{10}$Be production
 becomes significant at energies lower than 2 GeV/nucleon, while the overall production
 rate $R$ is mostly determined by protons.
The mean energy for $^{10}$Be production move in the course of a solar activity cycle, varying between
 1.5 GeV ($\Phi$=300 MV corresponding to recent solar cycle minima) and 4 GeV
 ($\Phi$=1000 MV corresponding to solar cycle maxima), which is somewhat larger than
 the value of 2 GeV used earlier (\cite{mccr01,usos02}).
This results from the fact that the magnitude of the solar cycle modulation of CR
 greatly increases towards lower energies, which decreases the effective energy
 responsible for the $^{10}$Be variations with respect to the mean energy.
The total $^{10}$Be production rate is shown in Fig.~\ref{Fig:map_pc} as a function of
 the heliospheric modulation strength, $\Phi$, and $P_c$.
The resulting $^{10}$Be concentration in polar ice is assumed to be proportional to the
 $^{10}$Be production rate,
 i.e., we ignore the details of the atmospheric transport and deposition of $^{10}$Be (\cite{masa99})
 since we are interested in much longer time scales.

We now invert the model by going through the sequence of Eq.~(\ref{Eq:model1}) in reverse order:
\begin{equation}
N\stackrel{(1')}{\longleftarrow} S \stackrel{(2')}{\longleftarrow} F_o\stackrel{(3')}{\longleftarrow}
\Phi\left(\stackrel{(4')}{\longleftarrow}CR\stackrel{(5')}{\longleftarrow}\right)\,^{10}{\rm Be}
\label{Eq:model2}
\end{equation}
Steps ($5^\prime$) and ($4^\prime$) are taken in one stride so that
 from a measured $^{10}$Be concentration in polar ice at
 a given location (given $P_c$),
 we determine the corresponding heliospheric modulation strength, $\Phi$, using
 the relationship shown in Fig.~\ref{Fig:map_pc}.
Steps ($1^\prime$)-($3^\prime$) are inversions of the corresponding steps (1)-(3) in the direct model.
In step ($3^\prime$), a power-law relation between $\Phi$ and the open magnetic flux, $F_o$, is used
 (\cite{usos02}):
\begin{equation}
F_o=0.023\ \Phi^{0.9},
\end{equation}
where $\Phi$ and $F_o$ are given in MV and $10^{14}\ {\rm Wb}$, respectively.

In step ($2^\prime$) we invert the following equation which connects
 the evolution of the open magnetic flux $F_o$
 with the source function $S$ (\cite{sola00}),
\begin{equation}
{dF_o\over dt}=S-{F_o\over\tau},
\label{Eq:S_non}
\end{equation}
where $\tau= 4\,$yr represents the characteristic decay time.
This equation includes differentiation, and when employing it to compute $S$
 from $F_o$, its inversion is stable
 only for smooth (noiseless) data series, for which  the time derivative does not fluctuate
 strongly from one data point to the next.
\begin{figure}
\resizebox{\hsize}{!}{\includegraphics{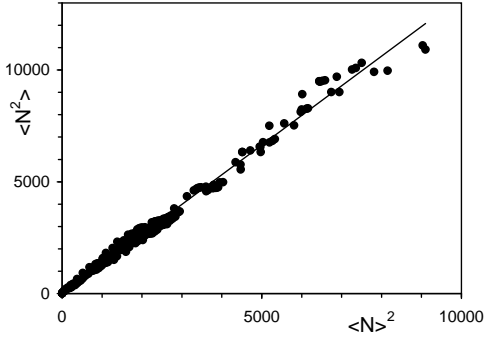}}
     \caption{Scatter plot of 11-year averaged $\langle N^2 \rangle$ vs.
      $\langle N\rangle ^2$ based upon the yearly group sunspot numbers for
      1700--2000.
      The regression line has a slope of $1.32\pm0.01$.}
     \label{Fig:r2}
\end{figure}
Real $^{10}$Be data are noisy and do not fulfill these conditions (see Fig.~\ref{Fig:data}).
Therefore, a
 reconstruction of yearly SN values from the yearly sampled Dye-3 Greenland $^{10}$Be
 record is extremely noisy.
However, averaging the yearly Greenland data by an 11-year running mean prior to reconstruction greatly
 reduces the noise and increases the stability of the procedure, although at the cost
 of losing information on variations shorter than a solar cycle.
Similarly, a direct reconstruction using the 8-year-sampled South Pole record is also
 somewhat noisy.
We therefore smooth the Antarctic data set using a 1-2-1 filter, which roughly
 corresponds to an 11-year smoothing.
In what follows we only consider the (effectively 11-year) averaged data series,
 so that Eq.~(\ref{Eq:S_non}) takes the form
  \begin{equation}
    \langle {dF_o\over dt}\rangle = \langle S\rangle
        - \frac{\langle{F_o}\rangle}{\tau}\ ,
\label{Eq:fi_av}
\end{equation}
where the angular brackets indicate temporal averages.
From the time sequence of averaged measured $^{10}$Be data we determine
 through steps ($3^\prime$)-($5^\prime$) the open flux values, $\langle{F_o}\rangle$.
The corresponding values of the average source function, $\langle{S}\rangle$,
 are obtained by replacing the time derivative
 in Eq.~(\ref{Eq:fi_av}) by a simple first-order finite difference.

The source function, $S$, is the following function of the sunspot number $N$
\begin{equation}
    S(N) = \alpha  \left( 24.35 + 22N - 0.061 N^2\right)\,,
\label{Eq:S}
\end{equation}
where $\alpha = 1.95\cdot10^{11}$ Wb/yr (\cite{sola00}).
In order to obtain the average sunspot number, $\langle{N}\rangle$, we have to
 to invert the time-averaged Eq.(\ref{Eq:S}) and to express $\langle
 N^2\rangle$ in terms of $\langle N\rangle$.
Unfortunately, there is no unique solution to this problem.
A regression analysis based upon the 11-year-averaged group sunspot number (GSN)
 record (\cite{hoyt98}) since 1700 leads to
 a factor $1.32\pm0.01$ (Fig.~\ref{Fig:r2}).
Therefore, we have used the relation $\langle N^2\rangle = 1.32  \langle N\rangle^2$
 in the inversion of Eq.(~\ref{Eq:S}).

\begin{figure}
\resizebox{\hsize}{!}{\includegraphics{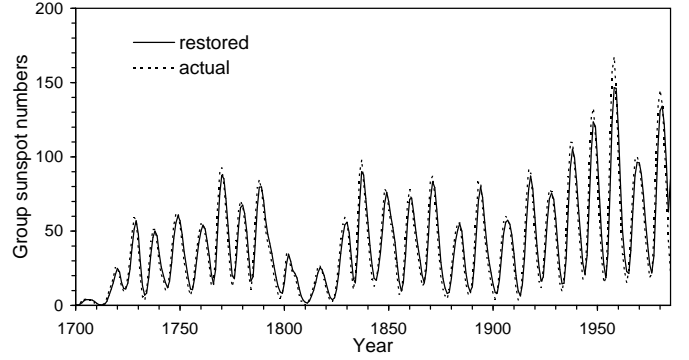}}
     \caption{Actual and restored yearly group sunspot numbers for 1700--1985 as a test
      of the consistency of the inversion model.}
     \label{Fig:test}
\end{figure}

In order to check the consistency of the full inversion procedure we have performed the following test.
Using the direct model (Eq.~\ref{Eq:model1}), we calculated, similar to (\cite{usos02}) but
 with the more realistic $^{10}$Be production described above,
 the expected $^{10}$Be concentration from the
 actual yearly group sunspot data (\cite{hoyt98}).
The calculated $^{10}$Be concentration was then used as an input for the inversion
 (Eq.~\ref{Eq:model2}), from which the corresponding restored group sunspot numbers were determined.
The result is shown in Fig.~\ref{Fig:test}.
Since the restored SN series is very close to the actual record (the RMS errors are
 $10.4$ for the yearly and $2.4$ for the 11-year averaged series), we conclude that the
 inversion procedure is consistent.
We emphasize that a nearly perfect reconstruction such as that in Fig.~\ref{Fig:test}
 is only possible for almost ideal data, with little noise, good sampling, etc.
The measured $^{10}$Be data do not fulfill these conditions nearly as well as the artificial
 record $^{10}$Be constructed from GSN, as discussed above.

\section{Reconstruction of sunspot numbers}
\begin{figure}
\resizebox{\hsize}{!}{\includegraphics{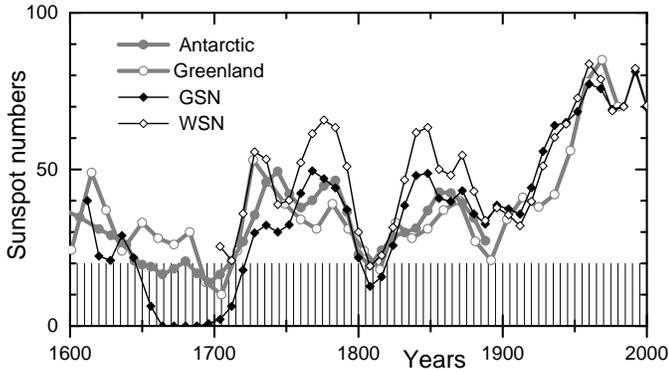}}
     \caption{11-year averaged sunspot numbers for the last 400 years: group sunspot
       numbers (GSN), Wolf sunspot numbers (WSN), SN reconstructed
        from Greenland $^{10}$Be and SN reconstructed from 1-2-1 filtered Antarctic $^{10}$Be data.
        The shaded area roughly indicates the  range of averaged SN for which
         the model is less reliable.}
     \label{Fig:SA1}
\end{figure}
As a first step, we have reconstructed the sunspot activity for the last 400 years
 from $^{10}$Be data using the model described above.
Since direct sunspot observations exist for this period, we can compare the
 reconstructed and measured SN.
The SN reconstructed from the two $^{10}$Be series are shown in Fig.~\ref{Fig:SA1},
 together with the measured Wolf and group
 sunspot numbers.
The agreement between the reconstructed SN series based on the 11-year averaged
 Greenland $^{10}$Be and the measured SN is quite good for the period 1700--1985,
 with an RMS discrepancy of 11 (the cross-correlation
 is $r=0.78^{+0.06}_{-0.10}$) for GSN and 12.5 ($r=0.74^{+0.07}_{-0.12}$) for the Wolf sunspot series.
The RMS deviation is 10 between the 1-2-1 filtered 8-year averaged GSN and SN
 reconstructed from the Antarctic $^{10}$Be series
 for 1700--1900 ($r=0.76^{+0.07}_{-0.12})$.

We note that the overall agreement between the reconstructed and measured SN is somewhat better
 for the group sunspot series than for the Wolf series.
For example, in 1770--1800 and 1830--1880, the reconstructed SN is very close to
 the GSN while the WSN is systematically higher.
The only exception is the period 1700--1750 when the reconstructed SN shows good agreement
 with the Wolf series while GSN is systematically lower (Fig.~\ref{Fig:SA1}).
This might imply that the group sunspot number values are somewhat too small during the first half
 of the $18^{th}$ century, in general agreement with the results of Letfus (2000).
We note that the two reconstructed SN series lie closer together
 than the GSN and WSN series in the common time interval,
 i.e., between 1700 and 1900.

In summary, we conclude that our method reconstructs the measured sunspot numbers
 reasonably well for the last 400 years.
The only exception is the Maunder minimum.
Whereas the actual sunspot number is close to zero (\cite{eddy76,ribe93,usos00})
 during this period, the reconstruction returns values between 10 and 25.
This is due to the fact that the model is not reliable during extended
 periods of very low solar activity, as was already pointed out (\cite{usos02}).
On the other hand, the SN is still well reconstructed
 during the Dalton minimum circa 1800.
Accordingly, we estimate that the reconstructed sunspot values are
 unreliable (overestimated) when SN$\leq 20$.
\begin{figure}
\resizebox{\hsize}{5.2cm}{\includegraphics{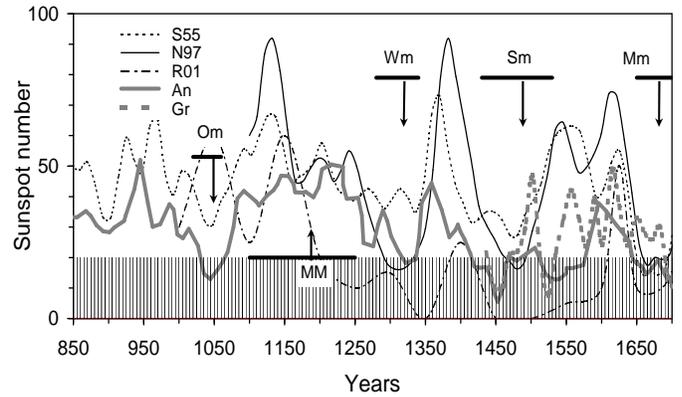}}
     \caption{Sunspot activity (11-year averaged) during the pre-instrumental era:
       earlier extrapolations by Schove (1955 - S55), Nagovitsyn (1997 - N97) and
        Rigozo et al. (2001 - R01), and our SN reconstructions from 1-2-1 filtered 8-year
         Antarctic (An) and from 11-year averaged Greenland (Gr) $^{10}$Be series.
       The shaded area indicates the range of averaged SN for which
         the applicability of the model is limited.
       Horizontal bars with arrows denote the known grand minima and maxima:
        Maunder minimum (Mm), Sp\"orer minimum (Sm), Wolf minimum (Wm),
        Oort minimum (Om) and the Medieval maximum (MM).}
     \label{Fig:SA2}
\end{figure}

In the next step, we have reconstructed the SN in the millennium time scale
 (Fig.~\ref{Fig:SA2}) using both the Greenland (since 1424) and the Antarctic
 (since 850) $^{10}$Be time series.
Great minima as deduced from  proxies like $^{14}$C, $^{10}$Be, etc.,
 also appear in our reconstruction, as does the Medieval maximum (\cite{stui89}).
We note that, because of the above-mentioned limitations of our model, the reconstructed SN,
 although significantly reduced during great
 minima, never reach zero.
While the SN reconstructions based upon the Antarctic and the Greenland data
 are rather consistent with each other after 1600 (Fig.~\ref{Fig:SA1}),
 they exhibit notable differences in 1480--1600.
The Antarctic series shows an extended minimum while the Greenland
 series exhibits  pronounced maxima in 1500 and in 1560.
Since this interval corresponds to the Little Ice Age, climatic differences
 between Greenland and Antarctica may well be the source.

Fig.~\ref{Fig:SA2} also shows that for the pre-instrumental era before 1600
our reconstruction differs  significantly from earlier results
 obtained by extrapolating multiharmonic representations of the
 measured SN (\cite{scho55,nago97,rigo01} - henceforth referred to as  S55, N97 and R01, respectively).
For example, the S55 and N97 series exhibit high maxima of activity around 1100,
 1400 and 1550--1600, which are either not present or significantly lower in our reconstruction.
On the other hand, R01 predicts a high maximum around 1050 when the Oort minimum, visible
 in all other series, is expected.
Throughout the whole
 period covered by Fig.~\ref{Fig:SA2}, our reconstruction shows much smaller variations of the
SN level  than the earlier series.

\section{Periodic features}

As discussed above, our reconstruction recovers the great minima and maxima of
 solar activity, which have been qualitatively deduced earlier.
We now  analyze the periodic features of the reconstructed SN series.
Since we deal with either 11-year averaged or 1-2-1 filtered 8-year sampled
 series, we will discuss only periods longer than 40 years.

\begin{figure}
\resizebox{\hsize}{!}{\includegraphics{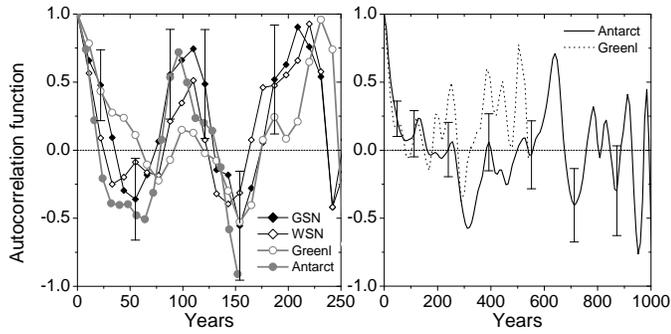}}
     \caption{Autocorrelation function of the measured SN (group, GSN, and Wolf, WSN)
      and the reconstructed (from Greenland and Antarctic $^{10}$Be data) sunspot series.
      Left panel: autocorrelation function for the period 1700--2000 (1700--1900
       for the Antarctic series; 1700--1985 for the Greenland series).
      Right panel: autocorrelation function for the period after 850 for the Antarctic
       series and after 1424 for the Greenland series.
      Error bars represent the 95\% confidence interval.}
     \label{Fig:auto}
\end{figure}

First we compare the periodicities in the reconstructed series with those in the
 measured SN after 1700.
In Fig.~\ref{Fig:auto} (left panel) we plot the autocorrelation function (ACF)
 for this period for the two reconstructed and two measured SN series.
The pattern is the same in all the four series, the main feature being
 a period of about  100 years, the Gleissberg secular cycle.
One can see that all the curves are rather close to each other, within the 95\% confidence
 interval, giving further support to the validity of our reconstruction method.
The right panel of Fig.~\ref{Fig:auto} shows the ACF for the last millennium for
 the reconstructions from the Greenland and Antarctic $^{10}$Be data.
Two dominant periodicities are present in the reconstructed data.
A cycle with an approximate period of 130 years is indicated by the consecutive peaks at 130,
 260, 400, 530, \dots years in both SN series.

Moreover, a cycle with a period of about 600 years visible as a deep
 minimum with the lag of about 300 years in both series and as a maximum
 at 600--650 years in the Antarctic series.

\begin{figure}
\resizebox{\hsize}{4.5cm}{\includegraphics{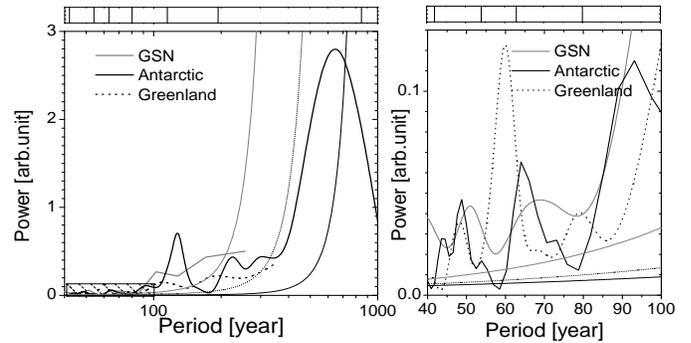}}
     \caption{Power spectra of the sunspot number as reconstructed
      from the Antarctic (850--1900) and Greenland (1425--1980) $^{10}$Be series and
      of the group sunspot number (GSN) series (1700--2000).
      The right panel is a zoom of the hatched rectangle in the left panel.
      Thin lines indicate the corresponding 95 \% confidence intervals.
      Bars in the upper panels represent the basic frequencies used by S55, N97 and R01 in
       their multi-harmonic models of sunspot activity.}
     \label{Fig:FFT}
\end{figure}

Fig.~\ref{Fig:FFT} presents the power spectra of the two reconstructed SN series
 and the GSN series.
The 95 \% confidence levels of the power spectrum (\cite{jenk69}) are shown by thin lines.
Note that the period after the Maunder minimum is too short to have a reliable power
 spectrum for the 8-year sampled Antarctic series, so that we have plotted
 the power spectrum using the whole data set.
The power spectra for the GSN and for the Antarctic data are in a fair agreement
in the period range from 40 to 100 years,
 revealing similar main periods: 52, 65--75 and 100 years in GSN series and 45--50,
 65 and 90--95 years in the Antarctic series.
The peak at 130 years, however, is not found in the GSN record.
The power spectrum of the Greenland series (1425--1980) shows a very
 strong peak at 60 years which is absent in the other two series.
Other significant peaks (50, 80 and 100--115 years) correspond roughly
 to similar peaks in both GSN and Antarctic power spectra.
Periods longer than half of the analyzed interval cannot be
 reliably determined as shown by thin lines in Fig. \ref{Fig:FFT}.
The power spectrum of the Antarctic series also exhibits a peak at about
 220 years probably related to the so-called de Vries or Suess cycle, (\cite{sues80,wagn01})
 and a prominent peak close to 600 years (\cite{sone90}).
The latter is in agreement with the autocorrelation function shown in Fig.~\ref{Fig:auto}.
At the top of Fig.~\ref{Fig:FFT} the periods used by previous investigators to
 extrapolate the SN are marked.
Not all of them agree with peaks in the power spectra of GSN and our reconstructions.
We also note that periods of different quasi-periodicities (such as, e.g., the Gleissberg cycle)
 seem to be fluctuating on the long-time scale.
This invalidates the earlier extrapolation of multi-harmonic representation of sunspot
 activity over extended time intervals.

\section{Geomagnetic variations}

\begin{figure}
\resizebox{7cm}{!}{\includegraphics{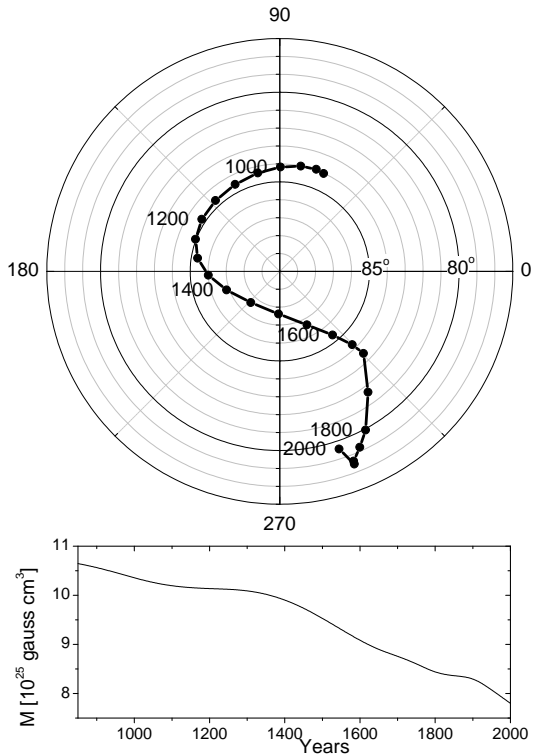}}
     \caption{Variation of the parameters of the geomagnetic dipole during 850-2000 AD.
     Migration of the dipole axis (top panel, in geographical coordinates)
      and change of the virtual dipole moment $M$ (lower panel). }
     \label{Fig:dipol}
\end{figure}

The above results were obtained assuming a local $^{10}$Be deposition and a constant
 rigidity cutoff of $P_c=0.5$ GV in Eq.~(\ref{Eq:prod}), which roughly corresponds
 to the atmospheric cutoff.
The geomagnetic field is known to change in time both in strength and orientation,
 which affects the $^{10}$Be production rate on time scales of centuries and longer
 (\cite{baum98}).
Fig. {\ref{Fig:dipol}} shows the variation of the geomagnetic field for the period
 850--2000 AD (see also Fig. 7 of \cite{hong98}).
The parameters of the geomagnetic field were taken from Hongre et al. (1998) for 850--1700 AD,
 from Bloxham \& Jackson (1992) for 1700--1900, and from the IGRF (International Geomagnetic Reference Field)
 model for 1900--2000 AD.
Using these data, we calculated the expected variation
 of the vertical geomagnetic cutoff, $P_c$, at the two $^{10}$Be sites using
 St{\o}rmer's equation (\cite{elsa56}):
\begin{equation}
P_c\propto M\cdot cos^4\Lambda,
\end{equation}
where $M$ is the virtual dipole moment and $\Lambda$ is the local geomagnetic latitude.
The calculated value of $P_c$ remains negligibly small (below 0.025 GV) for the South Pole
 during the whole interval studied here.
The geomagnetic latitude of the South Pole site remains above $80^{\circ}$, i.e. inside
 the polar cap where geomagnetic field lines are open.
Therefore, only the atmospheric cutoff was applied for the South Pole.
The value of $P_c$ for the Dye-3 Greenland site remains below 0.5 GV
 during the period of available data (since 1424) and,
 therefore, can be neglected as well.
At earlier times, however, the influence of the geomagnetic field becomes
 significant, e.g., the value of $P_c$ exceeds 1 GV for the Dye-3 site
 before the $14^{th}$ century.

\begin{figure}
\resizebox{\hsize}{!}{\includegraphics{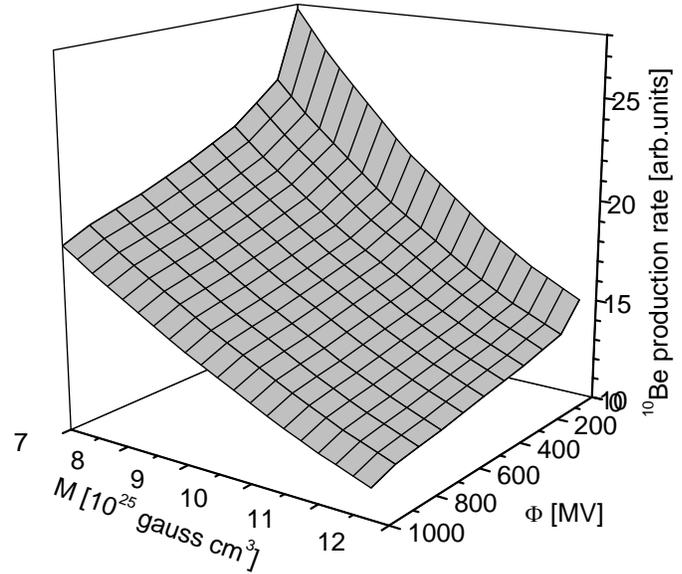}}
     \caption{Global production rate of $^{10}$Be in the atmosphere as a function of
      the virtual geomagnetic dipole moment, $M$, and the modulation strength, $\Phi$.}
     \label{Fig:map2}
\end{figure}

\begin{figure}
\resizebox{\hsize}{!}{\includegraphics{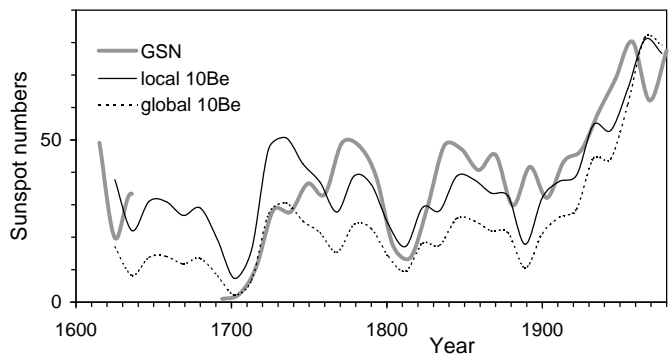}}
     \caption{11-year-averaged sunspot group sunspot number series (GSN) and
      the reconstructed SN for local and global $^{10}$Be deposition are
      represenetd by grey, solid, and dotted curves, respectively.}
     \label{Fig:SAgeom}
\end{figure}

We have also calculated the global $^{10}$Be production in the entire atmosphere
 as a function of the magnetic dipole moment and the modulation strength
 (Fig.~\ref{Fig:map2}, see also Fig.~2 in \cite{beer00}).
As an extreme case, we have performed a reconstruction of the SN assuming global mixing of $^{10}$Be before
 deposition, similar to $^{14}$C.
The result is shown in Fig.~\ref{Fig:SAgeom} and is compared with the reconstruction based upon
 local $^{10}$Be deposition and the GSN record.
Since the geomagnetic dipole was stronger in earlier times leading to a more effective shielding of
 the Earth, the same amount of globally distributed $^{10}$Be would require a stronger
 CR flux and, correspondingly, reduced sunspot activity.
Therefore, the SN level corresponding to the global $^{10}$Be model is systematically
 below the local model and also the GSN level.
This difference becomes increasingly larger for earlier times and the sunspot level
 goes systematically below 10 before 1600 in the global $^{10}$Be model.
This implies that the local deposition model is closer to reality than global mixing,
 although the truth probably lies in between.
Furthermore, the local $^{10}$Be model gives an upper limit to the level of sunspot activity
 since in the presence of atmospheric mixing the geomagnetic effects would require reduced
 activity levels at earlier times in order to retain consistency between the $^{10}$Be and GSN records.

\section{Conclusions}
The sunspot number record is not only the most widely used proxy of solar activity,
 it is also one of the longest running scientific data series.
Nonetheless, for many purposes it is still too short.
For example, both the statistical investigation of the solar dynamo and studies of a
 possible influences of solar activity on climate would greatly profit from
 a longer time series.

Since direct observations are sparse and not very reliable before the era of
 telescopic observations we have used an indirect proxy, the concentration
 of the cosmogenic $^{10}$Be isotope in the ice sheets in Greenland and
 Antarctica.
Unlike previous investigators, however, we have employed physical relations to
 convert the $^{10}$Be concentration into averaged
 sunspot numbers.
A comparison of directly observed SN with the reconstructed values for the last
 400 years confirms the reliability of the latter.
To enhance the reliability over longer times we have also considered the
 possible influence of geomagnetic field variations.
Taking the available data with a time resolution of the length of the solar
 cycle we have  reconstructed the SN back to 850 AD.

We have also performed a detailed analysis of periodicities in  the long-term
 solar activity record.
Both applied methods, autocorrelation function and power spectrum, reveal
that a number of strong periods in the directly measured sunspot series are also present in the
SN record reconstructed for the same period of time. Some differences between records are found, but they
are generally smaller than changes in dominant frequencies as different lengths of
the time series are considered.
In the higher frequency range, periods of about 50 and 65 years are visible in both measured and
 reconstructed series.
The secular Gleissberg cycle, which is roughly 100 years in the GSN series, is split into
 the dominant 120--130-year and subdominant 90--95-year cycles in the reconstructed series,
 the latter being more prominent during the last 3--4 centuries.
There is also evidence for an independent periodicity of about 220 years.
This may be related to the de Vries (also called Suess) cycle, earlier known in, e.g.,
 $^{14}$C isotope data.
A strong cycle with a period of roughly 600 year is apparent in the reconstructed series
based upon the Antarctic data.  Both the GSN record and the the SN series reconstructed from
the Greenland $^{10}$Be data are too short to exhibit this period, which is
possibly related to the 650-year periodicity in
 the $^{14}$C data (\cite{damo91}).

In conclusion, we have presented here a new reconstruction of solar activity on
 the millennium time scale based upon a description of the related physical processes.
The results will be the subject of further analysis.

\acknowledgements{We thank J\"urg Beer and Gennady Kovaltsov for useful comments as
 well as William Webber for providing us with the $^{10}$Be yield function.}

\end{document}